\documentclass[aps,pre,showkeys,showpacs,groupedaddress,amsmath,amssymb,floatfix,twocolumn]{revtex4-1}
\usepackage{amsmath}
\usepackage{amssymb}
\usepackage[utf8]{inputenc}
\usepackage{lmodern}
\usepackage[pdftex]{graphicx}
\usepackage[lofdepth,lotdepth]{subfig}
\usepackage{hyperref}
\usepackage{todonotes}
\usepackage{soul}
\usepackage{tikz}

\newcommand{\linkA}{\raisebox{2pt}{\tikz{\draw[-,blue,dashed,line width = 1.5pt](0,0) -- (4mm,0);}}}
\newcommand{\linkB}{\raisebox{2pt}{\tikz{\draw[-,red,dashdotted,line width = 1.5pt](0,0) -- (4mm,0);}}}

\newcommand{\mb}[1]{\mathbf{#1}}

\begin{document}

\title{Tweaking Synchronization by Connectivity Modifications}

\author{Paul~Schultz$^{1,2}$}
\email{pschultz@pik-potsdam.de}
\author{Thomas~Peron$^{1,3}$}
\author{Deniz~Eroglu$^{1,2}$}
\author{Thomas~Stemler$^{4}$}
\author{Gonzalo~Marcelo~Ram\'{i}rez~\'{A}vila$^{5}$}
\author{Francisco~A.~Rodrigues$^{6}$}
\author{J\"{u}rgen~Kurths$^{1,2,7,8}$}

\affiliation{$^1$Potsdam Institute for Climate Impact Research, P.O. Box 60 12 03, 14412 Potsdam, Germany\\
$^2$Department of Physics, Humboldt University of Berlin, Newtonstr. 15, 12489 Berlin, Germany\\
$^3$ Instituto de F\'{\i}sica de S\~{a}o Carlos, Universidade de S\~{a}o Paulo, S\~{a}o Carlos, S\~ao Paulo, Brazil \\
$^4$School of Mathematics and Statistics, The University of Western Australia, 35 Stirling Highway, Crawley, WA 6009, Australia\\
$^5$ Instituto de Investigaciones F\'{\i}sicas. Casilla 8635 Universidad Mayor de San Andr\'{e}s, La
Paz, Bolivia \\
$^6$ Instituto de Ci\^encias Matem\'{a}ticas e de Computa\c{c}ao, Universidade de S\~ao Paulo, CP 668, 13560-970 S\~ao Carlos, S\~ao Paulo, Brazil\\
$^7$Institute for Complex Systems and Mathematical Biology, University of Aberdeen, Aberdeen AB24 3UE, United Kingdom\\
$^8$Department of Control Theory, Nizhny Novgorod State University, Gagarin Avenue 23, 606950 Nizhny Novgorod, Russia}

\begin{abstract}
Natural and man--made networks often possess locally tree-like sub-structures. Taking such tree networks as our starting point, 
we show how the addition of links changes the synchronization properties of the network. We focus on two different methods 
of link addition. The first method adds single links that create cycles of a well-defined length. Following a topological approach we
introduce cycles of varying length and analyze how this feature, as well as the position in the network, alters the synchronous 
behaviour. We show that in particular short cycles can lead to a maximum change of the Laplacian's eigenvalue spectrum, 
dictating the synchronization properties of such networks. The second method connects a certain proportion of the initially
unconnected nodes. We simulate dynamical systems on these network topologies, with the nodes' local dynamics being either
a discrete or continuous. Here our main result is that a certain amount of additional links, with the relative position in the network
being crucial, can be beneficial to ensure stable synchronization.  
\end{abstract}


\maketitle

\section{Introduction}
\label{sec:introduction}
 
The study of dynamical processes on complex networks 
has been one of the most active fields within network science, where the ultimate goal is the precise evaluation of how network topology affects dynamics~\cite{barrat2008dynamical}. First numerical and analytical approaches to this problem were mainly concerned with the pure effect of the heterogeneity of the 
degree distribution on the overall network dynamics~\cite{barrat2008dynamical}. Great part of the interest in this subject has been
due to the detailed description of the topology of real systems that the network representation offers, a fact that motivates the search for a better understanding of their dynamical behavior as well. Early analytical developments were then mostly based on the configuration model~\cite{newman2001random}, which is able to construct networks with a given degree distribution.  Whilst offering a benchmark to study
dynamical processes in networks with any kind of degree
distribution, the networks constructed by such a model are generally locally tree-like, i.e. with a vanishing density of cycles~\cite{newman2010networks}. On the other hand, it is known that real networks in turn exhibit a much more sophisticated
topology, which encompasses features such as clustering, high-order loops, degree-degree correlations, community organization etc.~\cite{newman2010networks}. Thus, it is evident that studies based solely 
on the configuration model oversee topological characteristics that
are essential for a thorough analysis of the function of real networks. 

In order to overcome this limitation and take into account 
higher-order topological features, two strategies are commonly used, namely the adoption of networks constructed
through stochastic rewiring algorithms~\cite{kim2004performance,xulvi2004reshuffling,green2010large}
or variations of the traditional configuration model that
allow the creation of networks with tunable clustering or 
other kinds of subgraphs~\citep{newman2009random,karrer2010random,zlatic2012networks}.
In the former, one starts with a random tree-like network
and switches the edges according to some heuristics in order
to obtain a desirable network configuration, which is 
then used as a substrate for the dynamics under study. Although
this approach enables a precise control of a given network
property, as a function of which the dynamics can be analyzed, 
other properties are dramatically changed~\cite{kim2004performance,xulvi2004reshuffling,green2010large}. The latter makes the 
assessment of the isolated contribution of a particular topological property to the network dynamics unfeasible, since the results can be potentially influenced by spurious effects generated by the method. This limitation can be surpassed by extensions of the configuration model. However, depending on the subgraph structure modelled, the computational complexity quickly escalates, imposing further constraints on the analysis~\cite{karrer2010random,zlatic2012networks}. 

In this paper, we address the effect of particular structural patterns
found in real networks, namely cycles of different lengths,  by adopting
a different approach. In order to control their occurrence, and therefore evaluate their contribution to the network dynamics, here we consider tree networks with minimal link addition in a way that the number and the length of the cycles are precisely varied. Besides the thorough control of network structures being created, the minimal link addition adopted here opens the path for new strategies intended to enhance the stability of real networks. The reason for that resides in the fact that the creation and rewiring of links are usually costly tasks to be performed in these systems~\cite{barthelemy2011spatial}. 

A prominent example is that of power grids, whose proper functioning is vital for modern society. 
Since the connectivity pattern of the surroundings of a given dynamical unit strongly influences its stability 
~{\cite{Coletta2016,menck2014dead, Witthaut2012, Witthaut2016}}, it is crucial that the inclusion of new transmission lines is done in a way to ensure, or even enhance, the local and global stability of the network, while spending minimal amounts of resources. 
{Otherwise, counterintuitive dynamical effects like Braess' paradox lead to certain new 
links destabilizing synchronization by increasing the largest Lyapunov exponent, eventually changing it's sign ~\cite{Coletta2016, Witthaut2012}. This is deeply related to the
appearance of cycles in the network.}
Similar arguments also hold for other spatially embedded man-made, or
natural networks with constrained connectivity, for instance transportation or neuronal networks.

This paper contributes in this direction by quantifying the impact of cycles created under minimal link addition on the global network behaviour with a focus on stability. In general, however, we expect that our approach translates to a broad class of problems ranging from synchronization to  percolation, spreading processes \cite{Pikovsky2003, noh2004} or control theory \cite{preciado2010distributed}.
Regarding the network dynamics, we consider the nodes as identical oscillators operating in periodic or chaotic regimes 
in the paradigmatic cases of logistic maps and R\"ossler oscillators. 
We evaluate the stability of the synchronous regimes depending on the variation of the length of cycles in the network topology. 
By employing the Master Stability Function (MSF) formalism~\cite{barahona2002synchronization}, we map the problem into a 
spectral analysis of the Laplacian matrix. This spectral approach has also been successfully applied to reveal network-dependent coherence \cite{pereira2013}.  Recently, Pade and Pereira showed that link additions in directed networks can destabilize 
the synchronous regime \cite{pade2015}. 
{Furthermore, in the case of a removal of links or altered link weights, changes to the synchronous state and it's stability are 
also found to relate to the Laplacian spectrum ~\cite{Witthaut2016}, in leading order to the Fiedler eigenvector.
In this paper we study undirected networks and link addition, considering the whole stability 
interval of coupling values instead of only the lower boundary. }

Our results suggest that cycles of length four play a special role in 
network dynamics. More precisely, we find that the inclusion of links that 
create these motifs yields networks with higher synchronizability in 
comparison with cycles of different lengths. Furthermore, cycles of length 
three are found to have a weak effect on the Laplacian spectrum and, 
consequently, on the stability of the synchronized state. Interestingly, 
this peculiar innocuous effect of triadic connections on critical dynamical 
properties  has also been reported in other contexts~\cite{melnik2011unreasonable,yoon2011belief,herrero2015ising,peron2013synchronization,rodrigues2016kuramoto}.
 
This paper is organized as follows: In the next section we explain how we create different network topologies starting from a tree. We present two different methods (1) introducing only one cycle of a given length and (2) adding several random links. Moreover, we show in detail how different cycle lengths change the properties of the network that control the synchronizability. Thereafter, we study how the number of random links added to the starting tree impacts the synchronization of the network using two numerical models, namely the time discrete logistic map and the continuous R\"ossler system. Finally, we state the conclusions and perspectives.

\section{Network Topologies}
\subsection{Network Manipulation}

\begin{figure}[!tpb]
\centerline{\includegraphics[width=.9\linewidth]{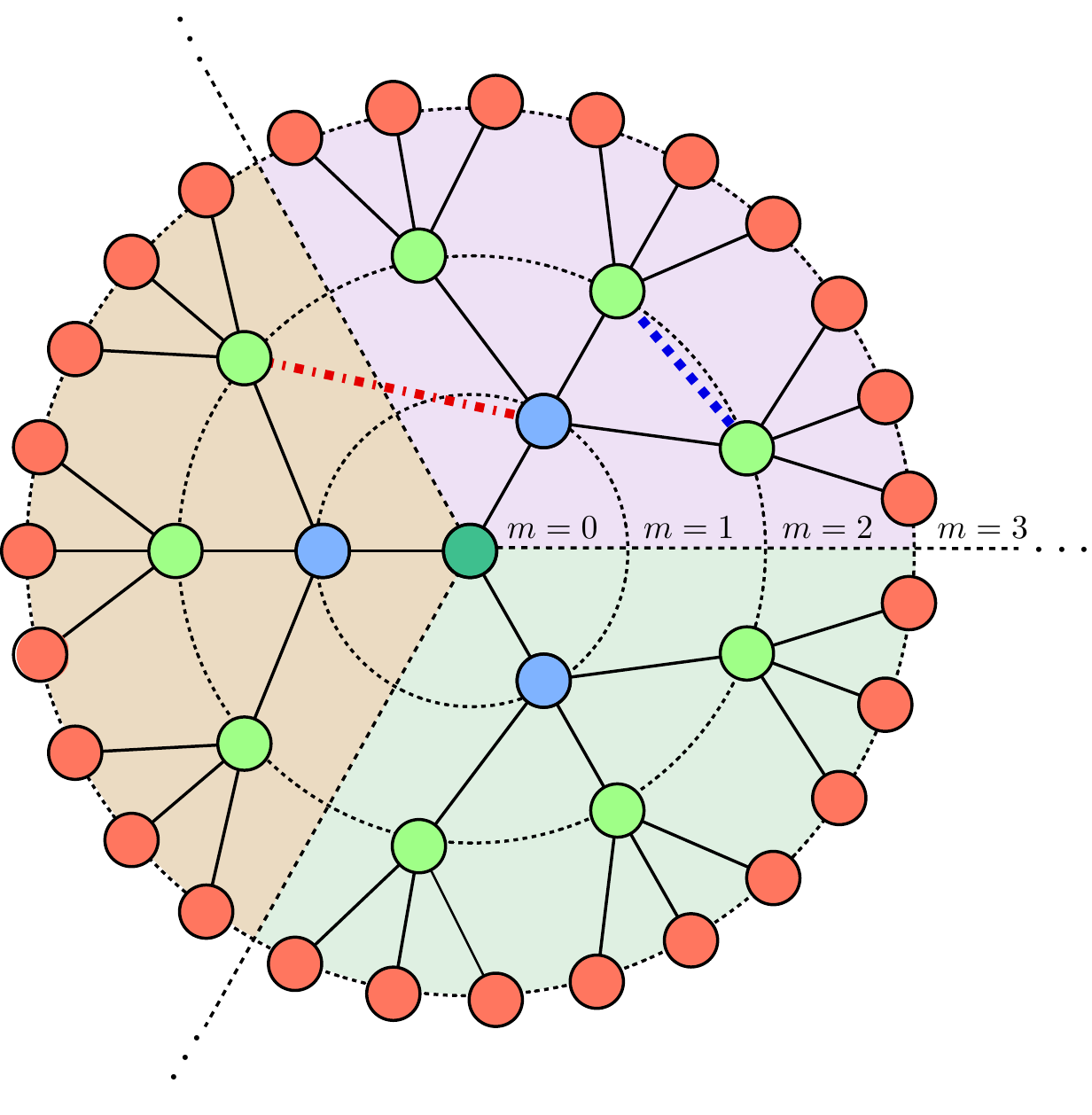}}
\caption{
(Color online) Sketch of a balanced tree network $G(m, 3)$. In the single link addition procedure, we distinguish two cases, 
namely connecting nodes in the same fundamental branch ("in", \protect\linkA) or in different 
("out", \protect\linkB), i.e. every path between the nodes before the addition contains the root node. }
\label{fig:network}
\end{figure}

The procedure of the network generation is as follows. As a starting point we consider an undirected balanced tree graph $G(m,n)$ where $m$ is the number of levels and $n$ is the branching number at each level. Therefore, the total number of nodes is given by $N = (n^{(m+1)}-1)/(n-1)$ and the total number of edges is $N-1$. Due to the fact that the total number of edges in a complete undirected graph is $N(N-1)/2$, the remaining unconnected number of pairs for the tree structure is given by $r = (N-2)(N-1)/2$. Notice that our starting network does not have any cycles. 

Having a highly structured -- tree-like -- network with known statistical characteristics allows us to control the network topology in two different ways. One is targeting the network topology directly by introducing cycles of a chosen length. The other way lets us to evaluate what happens if we control the sparseness of the network by randomly introducing a certain amount of additional edges. We are going to refer to the introduction of cycles as the \emph{single link addition}. Here the length of the cycle is the control parameter. The other method will be referred to as \emph{random link addition} and, in this case, the control parameter is the probability $p$ multiplying the number of unconnected pairs $r$. We want to point out that this network generation method is very similar to the ones used to generate \emph{small-world} structures \cite{newman1999}. However, the substantial difference is that our initial network is a balanced tree structure instead of a regular lattice.

What we aim to evaluate is how the different lengths of cycles respectively the choice of $p$-
values change the synchronization features of the networks. We consider the 
resulting network structure of our two methods as the adjacency matrix 
$A_{ij}$, with  $A_{ij} = 1$ if nodes $i$ and $j$ are connected, and 
$A_{ij}=0$ otherwise. The number of connections, the degree, of  node $i$ 
is given by $k_i=\sum_j A_{ij}$. Further, we define the Laplacian $\mathbf{L}$, 
$L_{ij} = \delta_{ij}k_i - A_{ij}$. Its eigenvalues $\lambda_i$ 
($\lambda_1=0\leq\lambda_2\leq\dots\leq\lambda_{\mathrm{max}}$)
play an important role in characterizing the 
synchronizability of the system and therefore measure comprehensively what we want 
to determine~\cite{barahona2002synchronization}, i.e. changes in the 
eigenvalue spectrum of our designed networks. In particular, we are 
interested in  the minimal changes or best cycle lengths that have maximum 
impact on the value of $\lambda_2$ or $\lambda_{\mathrm{max}}$ ($\lambda_1=0$ if 
the network is connected). The magnitude of the first non-trivial 
eigenvalue $\lambda_2$ is related to the onset of synchronization while 
the magnitude of the maximum eigenvalue $\lambda_{\mathrm{max}}$ is connected to the 
end of the synchronization interval~\cite{Boccaletti2006,newman2010networks}.

\subsection{Impact of Cycle Length}

\begin{figure}[h]
\centerline{\includegraphics[width=0.97\columnwidth]{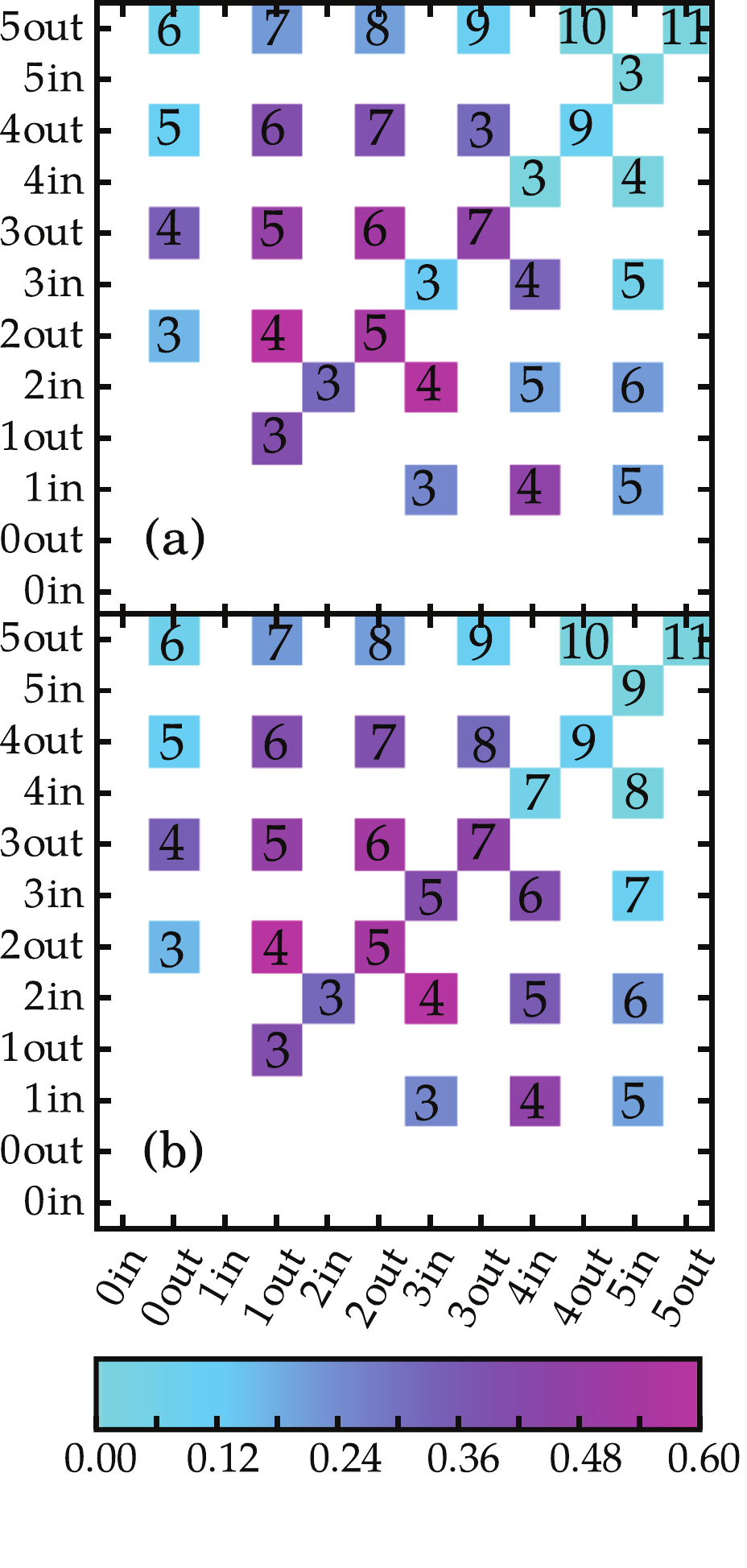}}
\caption{
(Color online)
{(a)} Minimal and  {(b)} maximal increase $\Delta\lambda_{\mathrm{max}}$ 
of the largest Laplacian eigenvalue $\lambda_{\mathrm{max}}$ for various wiring choices. For reference, the integers denominate 
the length of the {(a)} shortest and  {(b)} longest cycle {possible for each configuration}.
For the link classification "in" vs. "out" see Fig.~\ref{fig:network}.}
\label{fig:msf_part1}
\end{figure}

While we are going to study the random link addition later when 
focusing on network dynamics, we start by studying just the 
eigenvalue pair $\lambda_2$ and $\lambda_{\mathrm{max}}$ as a function of the 
cycle length after single link addition. Avoiding self-loops,
 the first cycle is a $3$--cycle, i.e. a node's neighbours in the network
are themselves connected. Using the illustration in 
Fig.~\ref{fig:network} of a balanced tree network $G(m,3)$ (only $m=3$ levels are shown) 
we see that introducing a $3$--cycle in this network means connecting two 
nodes $i$ and $j$ that lie on one of the circles highlighting the level 
number (see the dashed blue highlighted line in Fig. \ref{fig:network}). In addition, 
note that the branching on level $m=1$ of the original network is 
highlighted as the shading of the three areas. Ignoring the trivial level $m=1$, 
all of the 3--cycles that we can create lie within the same shaded area and 
do not break the symmetry of the network. This also means that they are 
linking two nodes within the same branch of the tree network (shaded areas 
with the same color in Fig.~\ref{fig:network}) and we call these links 
\emph{in-links}. On the other hand, if a link connects nodes belonging to 
different branches (e.g. the dashdotted red line in Fig. \ref{fig:network}),  we denominate 
it as an \emph{out-link}. In this case, the shortest cycle created by an out-link
is a $4$--cycle.

In Fig.~\ref{fig:msf_part1}a we can see how distinguishing between in- and out-links can be used as a tool to generate cycles of different lengths (shown for $G(5,3)$). Starting from the centre of the network, we number all nodes and their levels based on their distance. In this matrix-like plot we find that, for example, linking two nodes in the first level ($1$-out) creates a $3$--cycle. Similarly, a $3$--cycle can be created by linking from the second level ($2$-out) back to the centre ($0$-out). As we can see in $G(5,3)$, $11$--cycles are the longest possible cycles in this network. Note also that the average path length is changed by the introduction of different cycles. Especially short cycles will impact the average shortest path length while long cycles, e.g. out-links connecting leaf nodes, 
are likely to only have a small effect.


\begin{figure}[!tpb]
\includegraphics[width=\columnwidth]{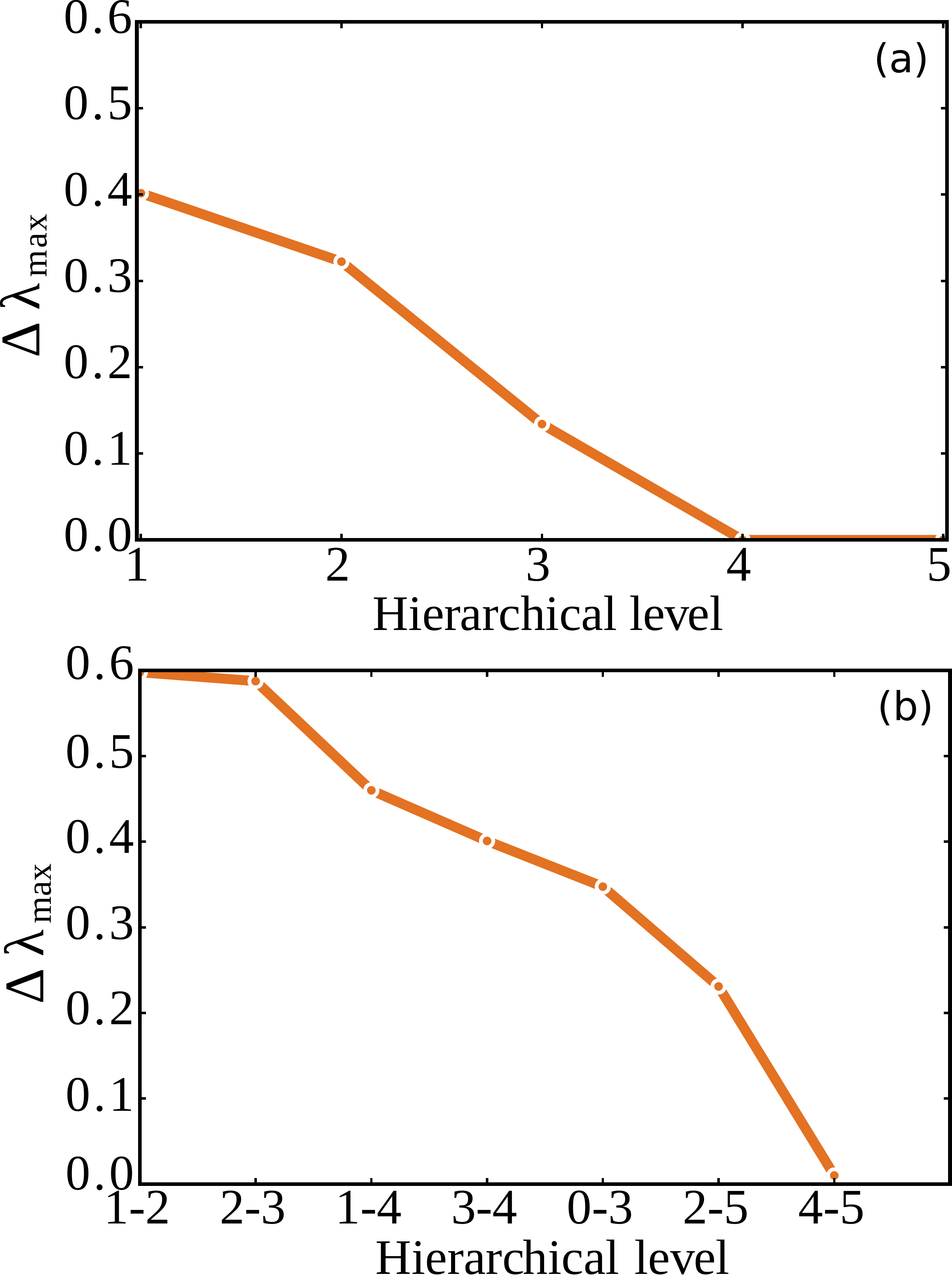}
\caption{
(Color online)
Increase of the maximum Laplacian eigenvalue $\lambda_{\mathrm{max}}$  for connections 
{(a)} on the same level creating cycles of length $3$ and {(b)} 
between different levels creating cycles of length $4$.}
\label{fig:sum_top}
\end{figure}

Since we are focusing on changes of the Laplacian's eigenvalues, it is worth pointing out that adding links in any undirected network always positively increases the magnitude of these eigenvalues. From a synchronization point of view we are, for instance,
 interested in the maximum possible change $\Delta\lambda_{\mathrm{max}}$ of the maximum Laplacian eigenvalue. 
In Fig.~\ref{fig:msf_part1} we show how much the different cycles change the magnitude of $\lambda_2$ and $\lambda_{\mathrm{max}}$. Again we give these changes in terms of \emph{in} and \emph{out} level connections. We want to highlight some of these changes. Firstly, we note that of the possible $3$--cycles the one with the strongest impact on the eigenvalues is on $m=1$, while $m\ge 2$ has less impact. In particular, connecting to neighbouring leaf nodes in level five does not change the upper limit of the 
spectrum while the largest sensitivity to topological changes is observed close to the root node. This tendency is repeated again in the $4$--cycles where the strongest impact is found when $2$--out is connected to a note on $m=1$ while already connections from $3$--out to $m= 2$ do not lead to such a large change. For the $3$ and $4$--cycle we summarize this topological effect in Fig.~\ref{fig:sum_top}. 

\begin{figure}[!tpb]
 \centering
 \includegraphics[width=\columnwidth]{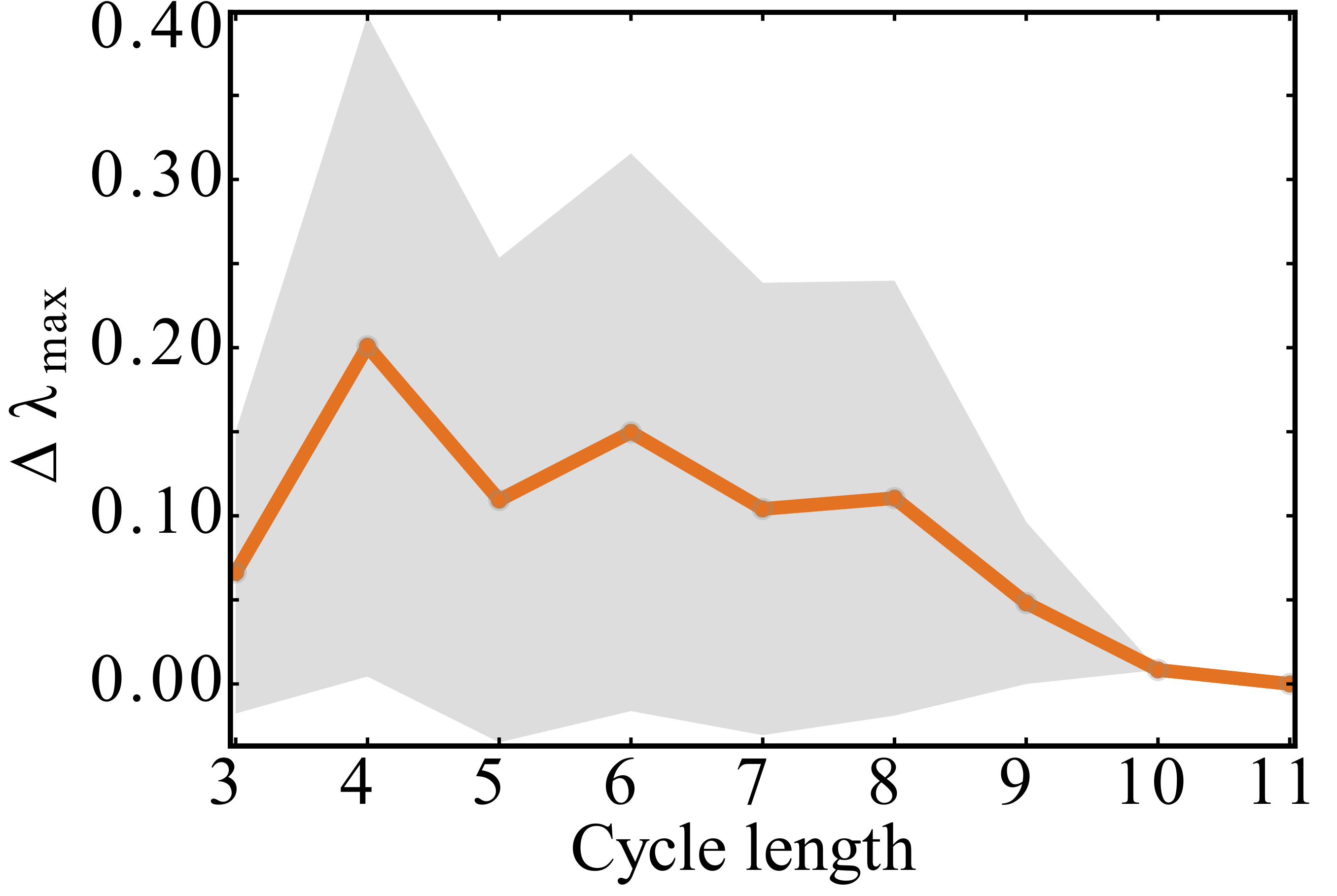}
\caption{
(Color online)
The average change of the maximum Laplacian eigenvalue
 is given as bullets while the shaded area indicates the width of the distribution between the minimal and the maximal change within the possible cycles in the network.}
\label{fig:cycle}
\end{figure}

If we ignore the additional information on where the cycle is within the network, it is a valid question to ask for the general impact of the cycle's length on changes in the eigenvalues. This information can be found in Fig.~\ref{fig:cycle} for the 
distribution of changes in the largest Laplacian eigenvalue of $G(5,3)$. It is not surprising that long cycles play almost no role in small-sized networks, while 4--cycles are dominating the change. Furthermore, 3--cycles induce changes in the spectrum comparable to long cycles, suggesting that such topological patterns have weak impact on network dynamics as well. This is somewhat remarkable, given how pervasive 3--cycles are in real-world networks~\cite{newman2010networks}. 
Noteworthy, it was verified (not shown here) that changes in $\Delta \lambda_{\max}$ due to the share of long cycles increase with the system size, but still $4$--cycles are dominating. $3$--cycles were consistently found to be responsible for lower changes in the spectrum than $4$--cycles. {This has also been addressed by Lodato et al. comparing the synchronizability of $4$-node subgraphs, quantified by the ratio $\lambda_{\mathrm{max}}/\lambda_2$ of the Laplacian eigenvalues~\cite{Lodato2007}. Their analysis showed that $4$-node motifs containing a single $3$-cycle have lower synchronizability than motifs with the same number of nodes and links containing a $4$-cycle. Interestingly, a single link addition that creates two $3$-cycles in a $4$-node motif does not increase its synchronizability. Therefore, their result suggests that networks with a higher number of $4$-cycles are more prone to exhibit a synchronized state.}

\begin{figure}[t]
\centerline{\includegraphics[width=\columnwidth]{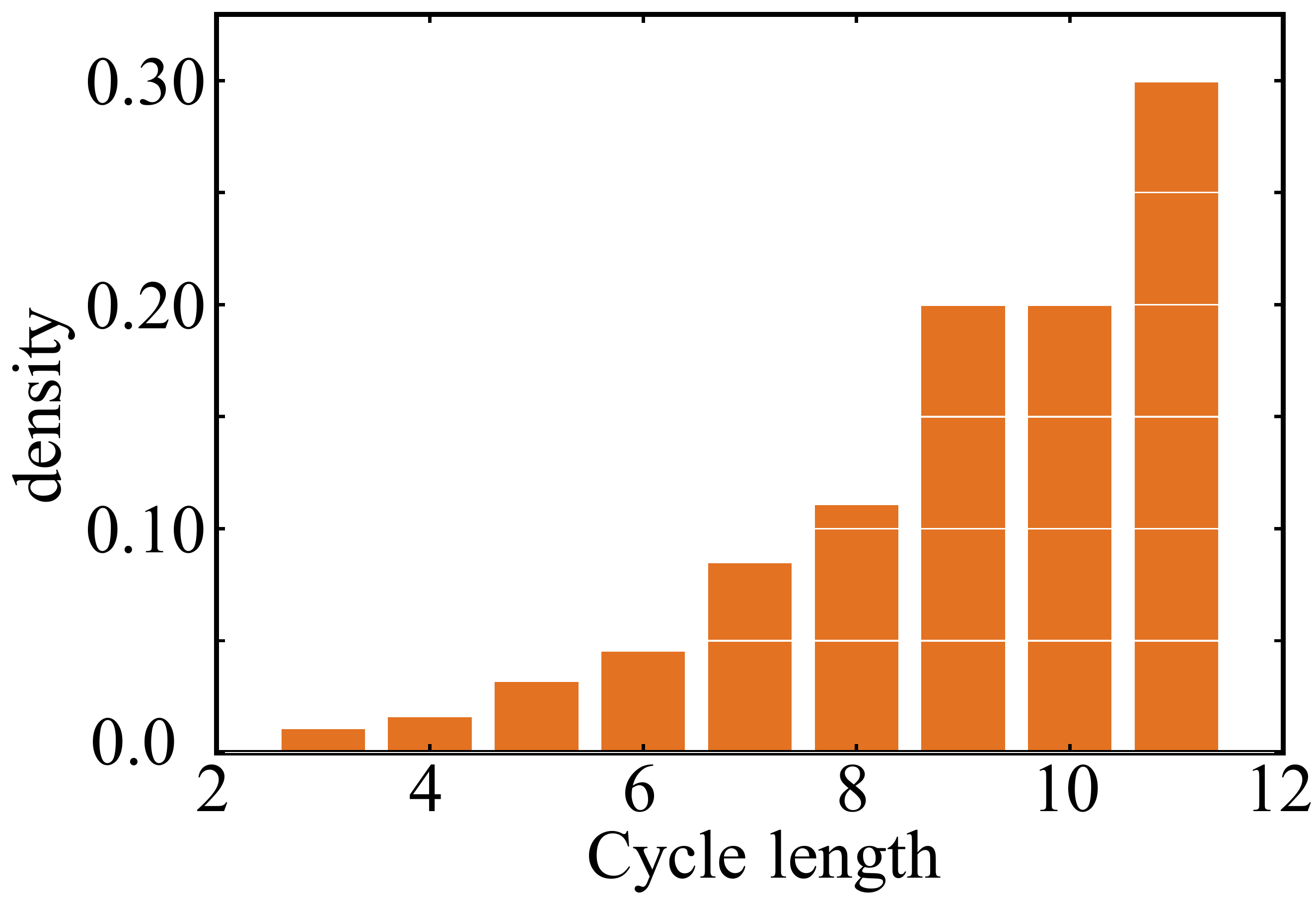}}
\caption{
(Color online)
Relative occurrence of cycle lengths when randomly connecting 
two nodes in a $G(5,3)$ balanced tree.}
\label{fig:hist}
\end{figure}

Concerning the random link addition, it is of great interest to estimate the probability for cycles of a particular length
being created. As we can see from the relative occurrence of cycle lengths plotted in Fig.~\ref{fig:hist} it is more likely that we choose longer cycles, i.e. on average a randomly chosen link would create a $9$--cycle. Therefore we can expect that large changes associated, for example, with some of the $4$--cycles will be less likely to dominate the eigenvalue spectrum, while the sum of changes resulting from several longer cycles will make up the main effect that changes the eigenvalue spectrum. 

\section{Master Stability Function} \label{sec:msf}

By means of the above mentioned methods, we obtain the adjacency matrices of the networks containing
our prototypical dynamic systems. We investigate two paragons of dynamical systems theory, namely R\"ossler oscillators, as an
example for an autonomous continuous system, and the well-established logistic map as a discrete one.
Our main tool allowing us to find relationships between the network spectral properties and the stability
of synchronous regimes of the considered dynamical systems is the master stability function (MSF) approach \citep{Pecora1998}.

\begin{figure}[t]
\centerline{\includegraphics[width=\columnwidth]{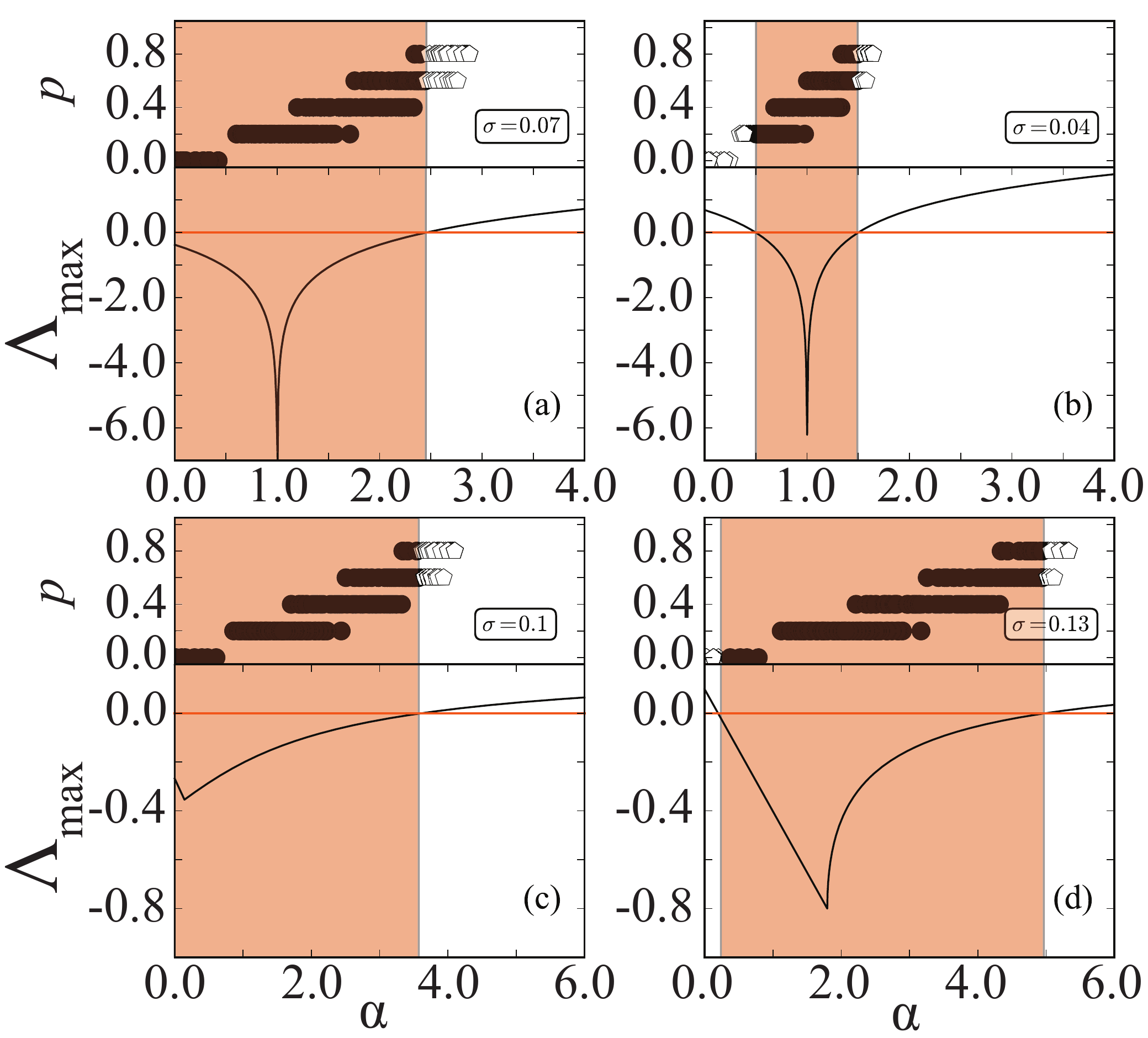}}
\caption{
(Color online)
Coupled logistic map in {(a)} periodic and {(b)} chaotic regimes; 
and R\"ossler oscillators in {(c)} periodic and {(d)} chaotic regimes. 
The upper plots depict the non-zero eigenvalues of the Laplacian $\mathbf{L}$ for different 
probability $p$ of link addition (see text for details). The {shaded} 
area denotes the existence of stable synchronous solutions. In all panels, the initial topology was $G(3,3)$.}
\label{fig:msf}
\end{figure}

To illustrate the method, we first assume a continuous (autonomous) dynamical system of, for instance, coupled oscillators described 
by a state vector $\mb{x}_i$ at site $i$ in a network:

\begin{equation}
\dot{\mb{x}}_i = \mb{F}(\mb{x}_i) + \sigma \sum\limits_{j=1}^N A_{ij} \mb{H}(\mb{x}_i,\mb{x}_j) \;,
\end{equation}
where $\mb{F}$ is the individual oscillator's dynamics, $\mb{H}$ the coupling function between coupled elements, $A_{ij}$ an element 
of  the network's adjacency matrix and $\sigma$ the overall coupling strength. To assess whether our networks allow for stable solutions of complete synchronization, i.e. $\mb{x}_i\equiv\mb{s}\; \forall i=1\dots N$, we need to obtain the MSF. For
identical oscillators this function is given by \citep{Pecora1998}:

\begin{equation}
\text{MSF}_{\mb{F},\mb{H}}(\alpha_i) = D\mb{F}(\mb{s}) - \alpha_i D\mb{H}(\mb{s})  \; ,
\end{equation}
where $\alpha_i = \sigma\lambda_i$ with Laplacian eigenvalues $\lambda_i$ 
and $D\mb{F}$ and $D\mb{H}$ are the Jacobian of the system and of the 
coupling function.
 If we assume for a moment that the argument $\alpha$ is a continuous (in 
general complex) variable, the real roots of the maximum Lyapunov exponent
$\Lambda_{\max}$ of $\text{MSF}_{\mb{F},\mb{H}}(\mb\alpha)$ determine the 
boundaries $\alpha_l$ and $\alpha_u$
of the stability interval. Note that for periodic systems, there is only one root $\alpha_u$ as $\alpha_l=0$.
 In order to have a stable solution,the coupling $\sigma$ needs to be chosen in a way that
  $\alpha_l < \alpha_i < \alpha_u\; \forall i=1\dots N$, for $\Lambda_{\max}$ 
  to be negative and hence to have an asymptotically stable synchronous regime (cf. Fig.~\ref{fig:msf}).
  
As discussed in the section before, the most crucial non-trivial Laplacian eigenvalues to be fit in the stability interval
are the second minimum $\lambda_2$ and the largest $\lambda_{\max}$ ones, since the rest of the eigenvalues are 
distributed between them. 

In the following, we apply this approach to the R\"ossler system as oscillatory units that are coupled through their $x$--coordinates:

\begin{align}
\dot{x}_i &= - y_i - z_i  + \sigma\sum\limits_{j=1}^N A_{ij} (x_j - x_i)\\
\nonumber\dot{y}_i &= x_i + a y_i \\
\nonumber\dot{z}_i &= b + z_i (x_i - c) \; .
\end{align}

Computing $\text{MSF}_{\mb{F},\mb{H}}(\alpha)$, we find that $D\mb{F}$ and $D\mb{H}$ are given by

\begin{align}
D\mb{F} &=
\begin{pmatrix}
0 & -1 & -1 \\ 1 & a & 0 \\ z^\ast & 0 & x^\ast-c
\end{pmatrix} \;\;
D\mb{H} &=
\begin{pmatrix}
1 & 0 & 0 \\ 0 & 0 & 0 \\ 0 & 0 & 0
\end{pmatrix} \; .
\end{align}

For our investigations we choose the parameters such that we have one set in the periodic regime ($a=0.1$, $b=0.2$, $c=5.7$)
and one in the chaotic regime ($a=0.2$, $b=0.2$, $c=5.7$, \citep{Roessler1976}). Notice that the MSF in general only yields a 
valid  linearization in a neighbourhood around a point $(x^*,y^*,z^*)^T$ from the synchronization
manifold which we determine from numerical simulations of the system. We obtain $ \alpha_u = 9.99884$ and 
$(\alpha_l, \alpha_u) = (0.198769, 4.99878)$ for the periodic and chaotic cases, respectively.

As a second application we use the time-discrete logistic map as an oscillator given by: 

\begin{align}
\label{eq:logmap}
x^i_{t+1} &= (1-\sigma)f(x^i_t) + \frac{\sigma}{k_i} \sum_j A_{ij} (f(x^i_t) - f(x^j_t)) \\
f(x_t^i) &= rx_t^i(1-x_t^i) \;,
\end{align}
where the local node dynamics $f(x^i_t)$  is the logistic map ($r \in [0,4]$) , $\sigma \in [0;1]$ 
is the coupling strength between the units and $A$ an adjacency matrix.

%

The logistic map is one-dimensional, therefore the Jacobian of Eq.~(\ref{eq:logmap}) is the derivative of $f$ with respect to $x_i$
and the associated MSF is given by 

\begin{align}
\label{eq:msf_logmap}
\text{MSF}_{f}(\alpha_i) = (1-\alpha_i)f'(x^i_t),
\end{align}
where $\alpha_i=\sigma \lambda_i$, $\lambda_i$ are the eigenvalues of the Laplacian as stated above. 

For our research we choose the control parameter of the logistic map, $r$, such that we have one periodic ($r = 3.83$, period 3) 
and one chaotic case ($r = 4.0$). For the periodic case, $\Lambda_{\max}$ again has only one root which is found as 
$\alpha_u = 2.45157$ (Fig.~\ref{fig:msf}). The roots for the chaotic local dynamics are found  to be $\alpha_l = 0.50038$ and
 $\alpha_u=1.49962$. 

\begin{figure}[t]
\centerline{\includegraphics[width=\columnwidth]{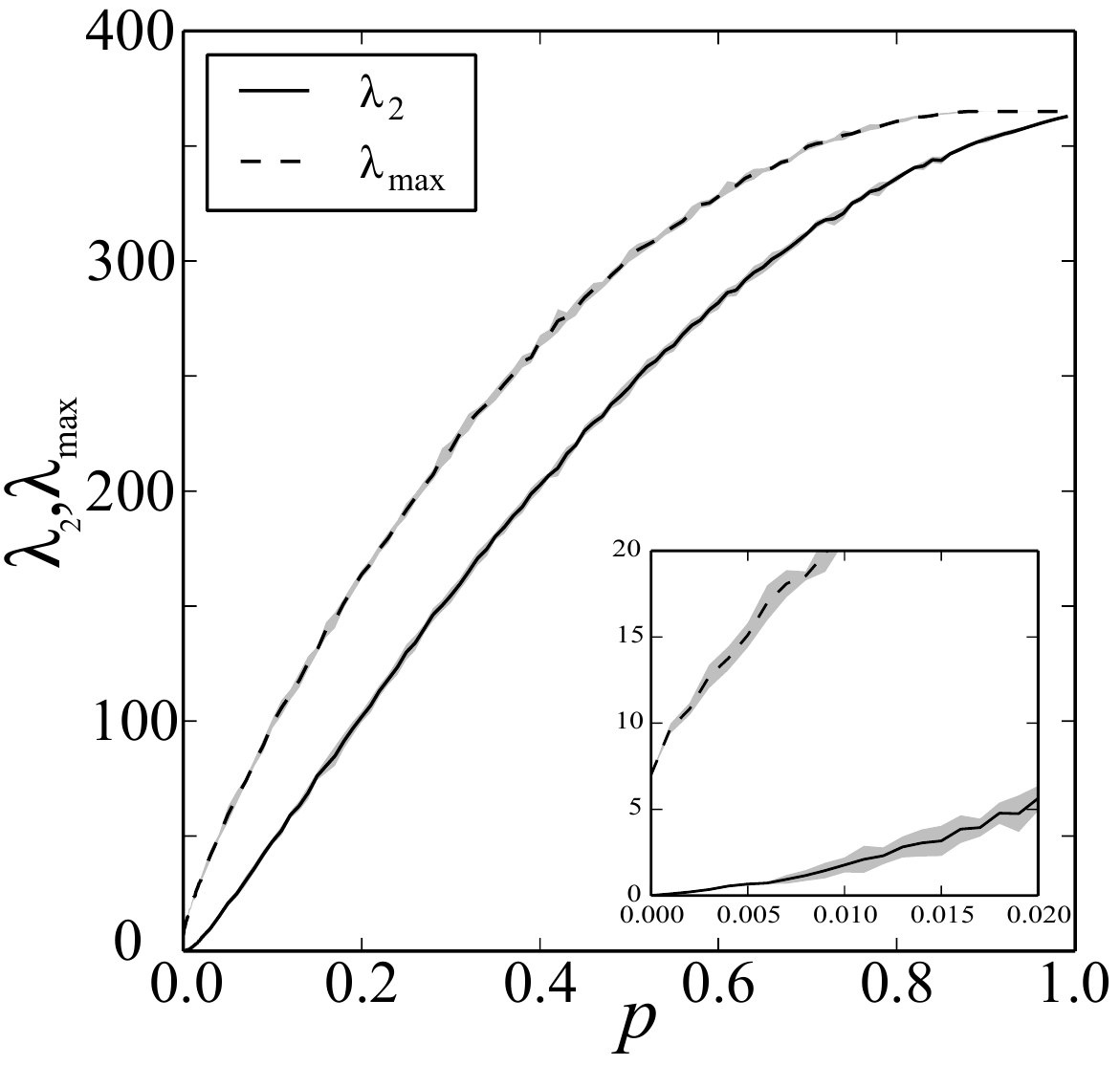}}
\caption{
(Color online)
{Pictured are the average eigenvalues $\lambda_2$ and $\lambda_{\max}$ for varying linkage probability $p$,
the shading indicates one standard deviation. The inset magnifies the regime of $0<p<0.02$. }
Note that the two eigenvalues $\lambda_2$ and $\lambda_{\max}$ bound the eigenvalue spectrum of the Laplacian.}
\label{fig:eig_rand}
\end{figure}

We demonstrate the stability of the coupled oscillators representing the nodes of our networks. We study the synchronization 
behaviour of two different node types, either the nodes' dynamics are given by  logistic map or the continuous R\"ossler system. 
Both systems are analyzed for two different dynamical regimes, periodic and chaotic. As we can see in Fig.~\ref{fig:eig_rand} the 
eigenvalues $\lambda_2$ and $\lambda_{\max}$ change with a different rate depending on the parameter $p$ which controls the 
random link addition. While the initial growth of $\lambda_{\max}$ dominates for low $p$-values, it seems to saturate for high 
values where we observe most of the change for $\lambda_2$. 

Hence, if just a few links are added at random locations, the initial growth of $\lambda_{\max}$ leads to a spreading of the
Laplacian spectrum, with the potential to cross the upper limit $\alpha_u$ of the synchronization interval. This can be 
counteracted by lowering the coupling $\sigma$, {i.e.} small changes in the network should be accommodated by reducing $\sigma$ to safely pertain in the synchronous interval.

In summary, Fig.~\ref{fig:msf} depicts the qualitative results for MSFs of the two test systems,
 the {shaded} areas denote the synchronous region. In the white regions the synchronous regime
is unstable.  As mentioned above, increasing $p$-values leads to an increase in edges and similarly the Laplacian's eigenvalues 
increase as well. 

i) The lower boundary for the periodic cases is $\alpha_l = 0$ (cf. Fig.~\ref{fig:msf}a,c), 
so that one only needs to consider the upper limit to ensure stable synchronization. 

ii) The lower boundary of the chaotic case is positive. 
Therefore both limits are needed to be considered (cf. Fig.~\ref{fig:msf}b,d). 

As we can see in all four cases,
there is a maximum amount of links that we can add before the system reaches the upper threshold of the synchronization region.  Clearly if we choose $p=0.8$, {some $\alpha$ values}  are no longer within the {shaded}  synchronization regions {in all panels of Fig.~\ref{fig:msf}}
(see the highlighted {pentagons} above each graph). More interesting is the situation for the oscillators being chaotic 
(cf. Fig.~\ref{fig:msf}b,d). We can see that without adding any links ($p=0$), the network does not support synchronization. Increasing the number of links leads to networks that can support synchronization, but again too many links will increase the eigenvalues by too much and we cannot find synchronization anymore. Comparing these results with the previous Fig.~\ref{fig:eig_rand} we see that $p$ needs to be big enough to increase $\lambda_2$ above the level where the lower bound is already in the synchronization region, while $\lambda_{\max}$ is still below the upper threshold of the band.

\section{Conclusion}
Since large natural or man-made networks are locally tree-like, we have focused on small trees as the starting point of constructing 
our network. We have seen that even including one edge can substantially alter the synchronization behaviour of the system 
and that short cycles connecting different levels of the tree have the largest impact on the eigenvalues of the Laplacian, especially if 
they are created close to the root node. We have highlighted a way to add cycles of defined length in our trees and therefore give the 
option to design networks having a particular synchronization behaviour.

Using the master stability framework we were able to analyze how random link addition alters the synchronization phenomenon using the logistic map as well as the R\"ossler oscillator. The most striking example, that we studied, is the synchronization behaviour of nodes having chaotic dynamics. While without additional links the networks are unable to synchronize the dynamics of the nodes, adding some links to the networks alters the global dynamics essentially and the systems can synchronize. Moreover we have found that adding too many links causes desynchronization.

In conclusion our work provides a method on how to optimize networks in such a way, that they can become synchronized in an
improved manner. Given that synchronization is of outermost importance in many networks, e.g. power grids~\cite{menck2014dead,rodrigues2016kuramoto,rohden2012self,motter2013spontaneous}, neuronal~\cite{rodrigues2016kuramoto,bullmore2009complex,arenas2008synchronization} and communication networks~\cite{arenas2008synchronization}, our findings may be used to increase the stability of the synchronization 
regime by adding short cycles.

\section*{Acknowledgement}

The authors wish to thank the Nesin foundation for an amazing working group activity in Nesin Math Village and wish to thank Tiago Pereira for fruitful discussions. PS and JK acknowledge gratefully the support of BMBF, CoNDyNet, FK. 03SF0472A. TP acknowledges FAPESP (No. 2012/22160-7 and No. 2015/02486-3) and IRTG 1740. DE acknowledge support by the Leibniz Association (WGL) under Grant No. SAW-2013-IZW-2. 


%
\end{document}